\documentclass[longauth]{aa}

\usepackage{graphicx}
\usepackage{txfonts}
\usepackage{float}
\usepackage{enumitem}
\usepackage{ctable}
\usepackage{easy-todo}

\begin{document} 

   \title{MINDS: Mid-infrared atomic and molecular hydrogen lines in the inner disk around a low-mass star}

    \author{Riccardo Franceschi\inst{1},
            Thomas Henning\inst{1},
            Beno\^{i}t Tabone\inst{2},
            Giulia Perotti\inst{1},
            Alessio Caratti o Garatti\inst{3,4},
            Giulio Bettoni\inst{5},
            Ewine F. van Dishoeck\inst{6,5},
            Inga Kamp\inst{7},
            Olivier Absil\inst{8},
            Manuel G\"udel\inst{9,10},
            G\"oran Olofsson\inst{11},
            L. B. F. M. Waters\inst{12,13},    
            Aditya M. Arabhavi\inst{14},
            Valentin Christiaens\inst{7},
            Danny Gasman\inst{15},
            Sierra L. Grant\inst{5},
            Hyerin Jang\inst{12},
            Donna Rodgers-Lee\inst{4},
            Matthias Samland\inst{1},
            Kamber Schwarz\inst{1},
            Milou Temmink\inst{6},
            %
            David Barrado\inst{16},
            Anthony Boccaletti\inst{17},
            Vincent Geers\inst{18},
            Pierre-Olivier Lagage\inst{19},
            Eric Pantin\inst{19},
            Tom P. Ray\inst{4},
            Silvia Scheithauer\inst{1},
            Bart Vandenbussche\inst{15},
            Gillian Wright\inst{18},
            %
        }
    
    \institute{Max Planck Institut f\"ur Astronomie (MPIA), K\"onigstuhl 17, 69117 Heidelberg, Germany
    \and Universit\'e Paris-Saclay, CNRS, Institut d’Astrophysique Spatiale, 91405, Orsay, France
    \and INAF – Osservatorio Astronomico di Capodimonte, Salita Moiariello 16, 80131 Napoli, Italy
    \and Dublin Institute for Advanced Studies, 31 Fitzwilliam Place, D02 XF86 Dublin, Ireland
    \and Max-Planck Institut f\"{u}r Extraterrestrische Physik (MPE), Giessenbachstr. 1, 85748, Garching, Germany
    \and Leiden Observatory, Leiden University, 2300 RA Leiden, The Netherlands
    \and Kapteyn Astronomical Institute, University of Groningen, The Netherlands
    \and STAR Institute, Universit\'e de Li\`ege, All\'ee du Six Ao\^ut 19c, 4000 Li\`ege, Belgium
    \and Dept. of Astrophysics, University of Vienna, T\"urkenschanzstr. 17, A-1180 Vienna, Austria
    \and ETH Z\"urich, Institute for Particle Physics and Astrophysics, Wolfgang-Pauli-Str. 27, 8093 Z\"urich, Switzerland
    \and Department of Astronomy, Stockholm University, AlbaNova University Center, 10691 Stockholm, Sweden
    \and Department of Astrophysics/IMAPP, Radboud University, PO Box 9010, 6500 GL Nijmegen, The Netherlands
    \and SRON Netherlands Institute for Space Research, Niels Bohrweg 4, NL-2333 CA Leiden, The Netherlands
    \and Kapteyn Astronomical Institute, Rijksuniversiteit Groningen, Postbus 800, 9700AV Groningen, The Netherlands
    \and Institute of Astronomy, KU Leuven, Celestijnenlaan 200D, 3001 Leuven, Belgium
    \and Centro de Astrobiolog\'ia (CAB), CSIC-INTA, ESAC Campus, Camino Bajo del Castillo s/n, 28692 Villanueva de la Ca\~nada, Madrid, Spain
    \and LESIA, Observatoire de Paris, Universit\'e PSL, CNRS, Sorbonne Universit\'e, Universit\'e de Paris, 5 place Jules Janssen, 92195 Meudon, France
    \and UK Astronomy Technology Centre, Royal Observatory Edinburgh, Blackford Hill, Edinburgh EH9 3HJ, UK
    \and Universit\'e Paris-Saclay, Universit\'e Paris Cit\'e, CEA, CNRS, AIM, F-91191 Gif-sur-Yvette, France
    }
 
  \abstract
   {Understanding the physical conditions of circumstellar material around young stars is crucial to star and planet formation studies. In particular, very low-mass stars ($M_\star < 0.2$~M$_\odot$) are interesting sources to characterize as they are known to host a diverse population of rocky planets. Molecular and atomic hydrogen lines can probe the properties of the circumstellar gas.}
   {This work aims to measure the mass accretion rate, the accretion luminosity, and more generally the physical conditions of the warm emitting gas in the inner disk of the very low-mass star 2MASS-J16053215-1933159. We investigate the source mid-infrared spectrum for atomic and molecular hydrogen line emission.
   }
   {We present the full James Webb Space Telescope (JWST) Mid-InfraRed Instrument (MIRI) Medium Resolution Spectrometer (MRS) spectrum of the protoplanetary disk around the very low-mass star 2MASS-J16053215-1933159 from the MINDS GTO program, previously shown to be abundant in hydrocarbon molecules. We analyzed the atomic and molecular hydrogen lines in this source by fitting one or multiple Gaussian profiles. We then built a rotational diagram for the H$_2$ lines to constrain the rotational temperature and column density of the gas. Finally, we compared the observed atomic line fluxes to predictions from two standard emission models.}
   {We identify five molecular hydrogen pure rotational lines and 16 atomic hydrogen recombination lines in the 5 to 20~$\mu$m spectral range. The spectrum indicates optically thin emission for both species. We use the molecular hydrogen lines to constrain the mass and temperature of the warm emitting gas. We derive a total gas mass of only $2.3 \times 10^{-5}$~M$_\mathrm{Jup}$ and a temperature of 635~K for the warm H$_2$ gas component located in the very inner disk ($r < 0.033$~au), which only accounts for a small fraction of the upper limit for the disk mass from continuum observations (0.2~M$_{Jup}$). The HI~(7-6) recombination line is used to measure the mass accretion rate ($4.0 \times 10^{-10}$~M$_\odot$~yr$^{-1}$) and luminosity ($3.1 \times 10^{-3}$~L$_\odot$) onto the central source. This line falls close to the HI~(11-8) line, however at the spectral resolution of JWST~MIRI we managed to measure both separately. Previous studies based on Spitzer have measured the combined flux of both lines to measure accretion rates. HI recombination lines can also be used to derive the physical properties of the gas using atomic recombination models. The model predictions of the atomic line relative intensities constrain the atomic hydrogen density to about $10^9-10^{10}$~cm$^{-3}$ and temperatures up to 5\,000~K.}
   {The JWST-MIRI MRS observations for the very low-mass star 2MASS-J16053215-1933159 reveal a large number of emission lines, many originating from atomic and molecular hydrogen because we are able to look into the disk warm molecular layer. Their analysis constrains the physical properties of the emitting gas and showcases the potential of JWST to deepen our understanding of the physical and chemical structure of protoplanetary disks.}
   

   \keywords{stars: pre-main sequence - stars: low-mass - accretion, accretion disks - protoplanetary disks - stars: variables: T Tauri -
techniques: spectroscopic
               }
    
   \authorrunning{R. Franceschi \inst{1}, Th. Henning\inst{1}}
              
   \titlerunning{Mid-infrared atomic and molecular hydrogen lines in the inner disk around a low-mass star}
   \authorrunning{Franceschi et al.}
   \maketitle

   \newcommand{\molhyd}{$\mathrm{H_2} \,$}
   \newcommand{\CO}[2]{$\mathrm{^{#1}C^{#2}O}$}
%

\section{Introduction}
\label{sec:intro}
While molecular hydrogen is the most abundant molecule in the universe, it is intrinsically hard to observe since it is a symmetric and homonuclear diatomic molecule \citep{Thi01, Carmona11, Trapman17}. Dipole ro-vibrational transitions are forbidden and only weak electric-quadrupole transitions are allowed. Molecular hydrogen emission can only be observed in high-temperature or high-luminosity environments. Its emission has been detected in several classical T~Tauri disks \citep{Beck08, Manara21, Gangi23}, coming from a warm layer above the optically thick dust emission \citep{Carmona11}. Atomic hydrogen lines are also often detected and originate from accretion flows \citep{Fang09, Hartmann16}. Given the challenges in observing hydrogen emission, protoplanetary disks are usually characterized through other observational signatures. Dust emission is more easily observed \citep{Andrews18}, and extensive modeling efforts can be found in the literature (see, e.g., \citealt{Testi14, Miotello22} for reviews). The disk gas structure remains less characterized, but with the development of observational facilities, it is now possible to study the gas through the emission of molecular tracers. For instance, the Atacama Large Millimeter/submillimeter Array (ALMA) observations of optically thin and optically thick CO isotopologues have been used to constrain the disk mass and temperature \citep{Miotello16, Miotello18, Zhang21, Calahan21, Schwarz16, Schwarz21, Pascucci23}. \citet{Yoshida22} measured the pressure broadening of CO line emission and derived the gas surface density in the inner disk of TW~Hya. \citet{Law22} and \citet{Paneque-Carreno23} used CO emission to study the disk vertical gas structure, while \citet{Lodato23} measured the disk mass from the perturbations on the CO gas orbit induced by the disk self-gravity in the disks around IM~Lup and GM~Aur. 

The HD molecule serves as a more direct indicator of H$_2$ since they are chemically very similar. The disk mass derived from HD measurements for TW~Hya is higher than the CO-based estimate by about two orders of magnitudes, likely due to the depletion of CO from the gas phase \citep{Bergin13, McClure16, Schwarz16, Kama20}. However, HD emission primarily originates from warm regions within the disk (around $T\approx 30-50$~K) \citep{Bergin13, Trapman17}, providing only lower limits to the gas mass. Additionally, HD has been detected only in TW~Hya \citep{Bergin13}, GM~Aur, and DM~Tau \citep{McClure16}, and no new HD observations have been conducted since the conclusion of the \textit{Herschel} mission.

Other emission lines are now commonly observed as well, each tracing different disk properties \citep{Oberg21}. HCO$^+$ and N$_2$H$^+$ trace the gas ionization in the molecular layer, a fundamental parameter to disk chemistry \citep{Cleeves15, Teague15, Aikawa21} and can also be used to measure gas depletion \citep{Smirnov-Pinchukov20, Anderson21, Trapman22}. 

The InfraRed Spectrograph (IRS) on the \textit{Spitzer} Space Telescope observed emission lines that can be used to characterize the gas surrounding young stellar objects (YSOs). For instance, a higher [Ne~II] luminosity in the inner disk originates either from stellar jets or from photoevaporative disks driven by stellar high-energy photons \citep{vanBoekel09, Lahuis07, Pascucci07, Gudel10, Baldovin-Saavedra11, Baldovin-Saavedra12}. \citet{Pascucci06} estimated the upper limit of gas masses in the inner disk of a sample of young solar-like stars from the non-detection of mid-infrared H$_2$ lines. They found that none of their systems have a mass greater than 0.05~M$_{Jup}$ within the first tens of au of the inner disk. Atomic hydrogen recombination lines are now commonly detected in the infrared spectra of YSOs and are used as an accretion tracer, as high HI luminosity is correlated with high accretion luminosities in the observed disks \citep[e.g.,][]{Antoniucci11, Alcala14, Antoniucci16, Beuther23}. In particular, \citet{Rigliaco15} detected the strongest of the Humphreys Series lines in the mid-infrared, the HI~(7-6) line ($12.37 \, \mu$m), and derived an empirical relation between the HI~(7-6) line luminosity and the accretion luminosity. 

This paper presents observations of the disk surrounding the very low-mass star 2MASS-J16053215-1933159 (referred to as J160532) using the James Webb Space Telescope (JWST) Mid-Infrared Instrument (MIRI) Medium Resolution Spectrometer (MRS). These observations were obtained as part of the MIRI mid-Infrared Disk Survey (MINDS). A comprehensive analysis of the MIRI data is detailed in \citet{Tabone23}, with a primary focus on hydrocarbon chemistry. The study reveals strong emission lines of C$_2$H$_2$, C$_4$H$_2$, and C$_6$H$_6$ while detecting no H$_2$O. This suggests that the inner disk of J160532 exhibits an abundance of hydrocarbons, indicating more active carbon chemistry in comparison to disks around higher-mass stars and a notably high C/O ratio.

The MIRI spectrum of this source shows numerous atomic and molecular hydrogen emission lines that we investigate in this paper. We used the HI recombination lines to measure the source accretion rate and the atomic gas properties and the H$_2$ emission to constrain the gas rotational temperature and column density.

\begin{figure*}
    \centering
    \includegraphics[width=\linewidth]{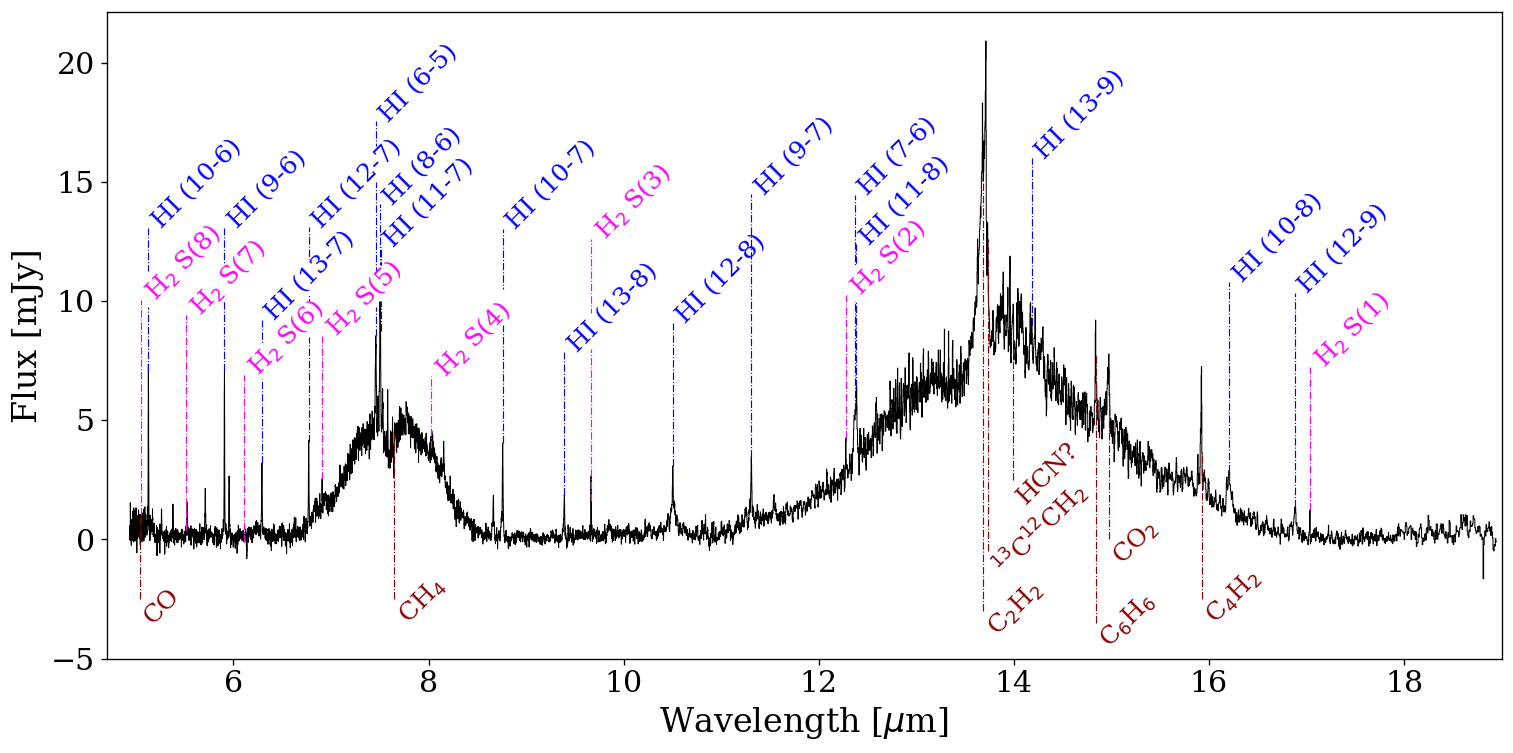}
    \caption{JWST MIRI-MRS continuum-subtracted spectrum of J160532. We identify several molecular (pink) and atomic hydrogen (blue) transitions, highlighted in the figure. A study of other molecular species (dark red) can be found in \citet{Tabone23}}
    \label{fig: spectrum}
\end{figure*}

\section{Observations}
\label{sec: observations}
The target source, J160532, is an M dwarf star, the most common class of stars found in our Galaxy, and frequently hosting exoplanets \citep{Dressing15, Sabotta21}. Constraining the physical properties of disks around these stars provides important information about planet formation models. The star is located in the Upper Scorpius star-forming region, with a distance of $152 \pm 1$~pc \citep{Gaia} and an age of $2.6 \pm 1.6$~Myr \citep{Miret-Roig22}. The source is an M4.75 spectral type star, with mass $M_\star = 0.14$~M$_\odot$ and luminosity $L_\star = 0.04$~L$_\odot$ \citep{Pascucci13, Luhman18, Carpenter14}. The spectral energy distribution (SED) of the source indicates the presence of a protoplanetary disk and the non-detection of millimeter continuum emission indicates a dust mass less than 0.75~M$_\mathrm{Earth}$, corresponding to a total gas and dust mass of less than 0.2~M$_\mathrm{Jup}$ for a standard gas-to-dust ratio of 100 \citep{Barenfeld16}.

The data analyzed in this paper are introduced in \citet{Tabone23}, where details about the observations, data reduction, and analysis are presented. Below we briefly summarize the main aspects. The data were observed by the JWST~MIRI instrument \citep{Rieke15, Wright15, Labiano21, Wright23} in Medium Resolution Spectroscopy (MRS) mode \citep{Wells15, Argyriou23} on 2022~August~1, for a total exposure time of 2.22~hours. These data were observed as part of the Cycle~1 Guaranteed Time Observation (GTO) program 1282, the MINDS program (PI: Thomas Henning). The details of the data reduction of the uncalibrated data can be found in \citet{Tabone23}. The spectral resolving power $R$ ranges from 3500 to 1500 from the shortest to the longest wavelengths, equivalent to a velocity resolution of about 90~km~s$^{-1}$ to 200~km~s$^{-1}$. The spectrum has been corrected for the source small radial velocity -1.98~km~s$^{-1}$ \citep{Jonsson20}. The extinction for this source is particularly low, and no correction was needed for these data. To better analyze the spectral features, we performed a low-order fit of the continuum and of the two broad bumps at 7.7~$\mu$m and 13.7~$\mu$m, corresponding to optically thick C$_2$H$_2$ emission (see \citealt{Tabone23} for a detailed description of the procedure). The resulting spectrum is shown in Fig.\ref{fig: spectrum}.

\section{Analysis and results}
\subsection{Molecular hydrogen}
\label{sec: molecular hydrogen}
\begin{figure*}
    \centering
    \includegraphics[width=\linewidth]{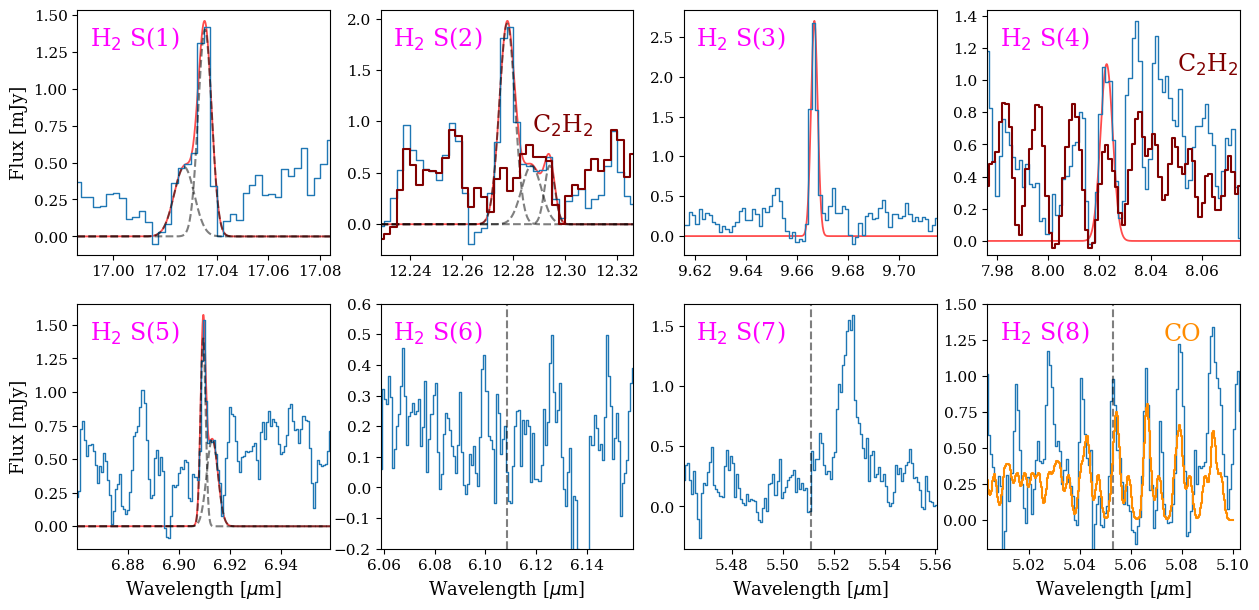}
    \caption{Molecular hydrogen pure rotational lines within the spectral range. The red profiles show line fitting using one or multiple Gaussian functions (dashed lines) in the case of line blending with other spectral features. The rest wavelength of the undetected lines is indicated by a slashed vertical line. In dark red we show the C$_2$H$_2$ rescaled model from \citep{Tabone23} to show the features location. In orange, we show an eye-fit of a CO slab model. The $S(2)$ line is blended with a C$_2$H$_2$ line. The $S(6)$ and $S(7)$ transitions fall in a crowded spectral region, and could not be identified. The $S(8)$ transition is found in a region populated by several CO~$v=1-0$ P-branch lines, and cannot be identified either.}
    \label{fig:h2}
\end{figure*}

As reported in \citet{Tabone23}, the spectrum shows HI, H$_2$, and molecular (C$_2$H$_2$, C$_4$H$_2$, CO$_2$ among many others) features. A study of these molecular components of the spectrum is also included. Here we focused on the HI and H$_2$ emission lines, reported in Fig.~\ref{fig: spectrum}. Thanks to the high resolving power and sensitivity of the MIRI-MRS instrument, we detected five pure rotational $\nu=0-0$ transitions of H$_2$, while the typically fainter rotational transitions with higher vibrational states remain undetected (Fig.~\ref{fig:h2}). The emitting region is unresolved, with no evidence of extended emission, meaning that it is coming from the inner, warm region of the disk. These lines are expected to be optically thin due to the small transition probabilities, and they can be used to constrain the temperature and column density of the emitting gas.

To accurately measure the line fluxes, we performed a linear fit of the continuum in a wavelength window of 0.03~$\mu$m centered on each line. The error in the flux density is the standard deviation on a nearby spectral segment that does not contain spectral features. The H$_2$ lines were fitted using one or multiple Gaussian profiles when the line showed evidence of blending with other spectral features to remove their contribution. This was done using the Levenberg-Marquardt least-squares minimization. The H$_2$ line flux was then calculated by integrating the main Gaussian component. In our calculations, the width of the Gaussian profiles was left as a free parameter. However, when working at the spectral resolution scale, one could also fix the width of the Gaussian profiles at the local spectral resolution element. We found that both methods give very similar results, within our error estimates, and we opted to leave the Gaussian width as a free parameter. The $\nu=0-0$~$S(6)$, $S(7)$, and $S(8)$ lines are in a spectral region rich in other spectral features, and their profiles could not be extracted. In particular, the $S(8)$ line falls between 4.9 and 5.1 $\mu$m, where several CO~$\nu=1-0$ P-branch lines are found, indicative of high-temperature gas ($T > 1000$~K) \citep{Tabone23}. The $S(1)$ and $S(4)$ are partially blended to C$_2$H$_2$ emission lines, identified in a previous analysis \citep{Tabone23}. However, the C$_2$H$_2$ emission contributes only a small amount to the total, and they produce results consistent with those from the other H$_2$ rotational lines. The line profiles and the best-fit profiles are shown in Fig.~\ref{fig:h2}, and the line parameters are reported in Tab. \ref{tab: H2}.

\begin{table}[]
    \caption{Measured parameters of the identified pure rotational molecular hydrogen lines.}
    \label{tab: H2}
    \centering
    \begin{tabular}{c c c}
        \hline
        \hline
        Transition & Line Center & Integrated Flux\\
         & [$\mu$m] & $10^{-13}$ [erg~s$^{-1}$~cm$^{-2}$]\\
        \hline
        $S(1)$ & 17.035 & $0.9 \pm 0.2$\\
        $S(2)$ & 12.279 & $2.4 \pm 0.4$\\
        $S(3)$ & 9.665 & $2.8 \pm 0.2$\\
        $S(4)$ & 8.025 & $3.1 \pm 0.9$\\
        $S(5)$ & 6.910 & $2.2 \pm 0.7$\\
        \hline
    \end{tabular}

\end{table}

We then used the measured fluxes to build a rotational diagram \citep{Goldsmith99} to estimate the emitting gas temperature and column density, assuming that the H$_2$ emission is in local thermodynamical equilibrium (LTE). This assumption is justified by the low critical density of H$_2$ emission, which is the typical minimal density required to populate the energy levels by collisions binging them to LTE. The critical densities for H$_2$ rotational levels are orders of magnitude lower than for CO, the next most abundant molecule, and lower than the typical densities in the warm molecular layers. The difference in critical densities is due to the lower Einstein coefficient probability of rotational H$_2$ transitions compared to CO rotational transitions. The low critical density for H$_2$ rotational transitions, therefore, ensures LTE emission.

Assuming a Boltzmann distribution, for optically thin emission the observed fluxes $F_J$ must satisfy the relations:

\begin{equation}
    \label{eq:boltzmann}
    \frac{4 \pi F_\mathrm{J} }{h  c \nu_\mathrm{J} \Omega g_\mathrm{J} A_\mathrm{J}} = \frac{N_\mathrm{tot}}{Q_\mathrm{rot}} e^{-E_\mathrm{J} / k_\mathrm{B} T_\mathrm{rot}}
\end{equation}

\noindent where $h$ is the Planck constant, $k_\mathrm{B}$ the Boltzmann constant, $c$ is the speed of light, $\nu_J$ is the line frequency, $\Omega = \pi(r_\mathrm{em} / d)^2$ is the emitting area solid angle, $A_{J}$ the Einstein coefficient, $g_J = 2J +1$ is the statistical weight of the transition, $N_\mathrm{tot}$ is the molecule column density, $Q(T_\mathrm{rot})$ is the partition function, $E_J$ is the transition energy, and $T_\mathrm{rot}$ is the rotation temperature. The ortho-to-para ratio is assumed to be 3 in LTE. Assuming optically thin emission at LTE with temperature $T_\mathrm{rot}$, we estimated the column density and temperature of the gas:

\begin{equation}
    \label{eq:diagram}
    \ln \left( \frac{4 \pi F_\mathrm{J}}{h c \nu_\mathrm{J} g_\mathrm{J} A_\mathrm{J}} \right) = -\frac{E_\mathrm{J}}{k_\mathrm{B} T_\mathrm{rot}} - \ln \left( \frac{Q_\mathrm{rot}}{N_\mathrm{tot} \Omega} \right).
\end{equation}

\noindent This expression allows the construction of a rotational diagram. In this graphical representation, the variables are $x = E_\mathrm{J} / k_\mathrm{B}$ and $y = \log \left( N_\mathrm{J} / g_\mathrm{J}\right)$, where $N_\mathrm{J} = 4 \pi F_\mathrm{J} / h c \nu_\mathrm{J} A_\mathrm{J}$. The second term on the right-hand side is treated as a constant. When plotted, this diagram is expected to produce a straight line with a slope of $-1/T_\mathrm{rot}$ and an intercept that provides the mass of the emitting gas, under the assumption of a specific emission area. 

\begin{figure}
    \centering
    \includegraphics[width=\linewidth]{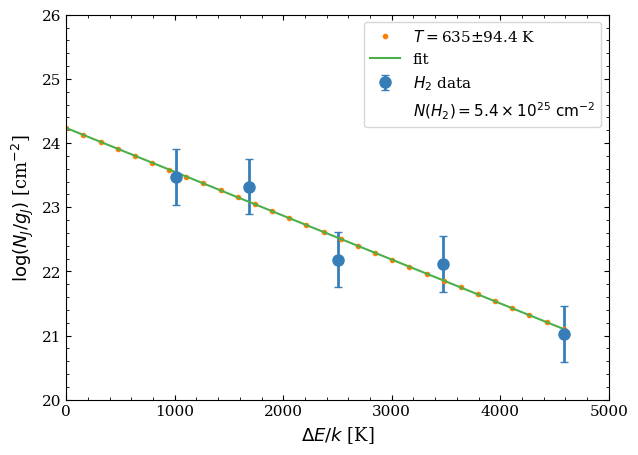}
    \caption{Rotation diagram of H$_2$ rotational transitions. A linear fit was performed on the measured fluxes, and the diagram shows a single component for the emitting gas with a temperature of about 635~K and mass $2.3 \times 10^{-5}$~M$_\mathrm{Jup}$.}
    \label{fig: diagram}
\end{figure}

The resulting rotation diagram is shown in Fig.~\ref{fig: diagram}. The emitting area can be estimated from an optically thick line. We assumed the H$_2$ emission to have the same emitting radius $r_\mathrm{em}$ as the C$_2$H$_2$ emission, 0.033~au \citep{Tabone23}. Indeed, the emitting radius of the H$_2$ emission cannot be directly estimated due to its optically thin nature, which leads to a degeneracy between column density and temperature of unresolved H$_2$ emission. The C$_2$H$_2$ emission, on the other hand, is optically thick and independent from column density, allowing us to estimate its emitting radius. Since the emission of C$_2$H$_2$ and H$_2$ are excited by similar physical conditions, it is reasonable to assume that the C$_2$H$_2$ and H$_2$ emission originate from the same disk region, and have the same emitting radius. Using this estimate for the emitting radius, we found a temperature of $635 \pm 94$~K and a column density of $190 \pm 110$~g~cm$^{-2}$, corresponding to a total mass of warm H$_2$ gas of $(2.3 \pm 1.3) \times 10^{-5}$~M$_\mathrm{Jup}$. 

We tested the consistency of these results by running a slab model for the H$_2$ emission. In this model, the line emission is assumed to originate from a gas layer of temperature T, column density of N, and emitting radius r, similarly as in \citet{Tabone23}. Using our temperature and mass estimates and our assumed emitting radius we obtained results consistent with the observed line fluxes. Moreover, the predicted fluxes for the S(6), S(7), and S(8) lines fall below the noise level of our observations, consistent with their non-detection in the data.

The linear trend observed in the rotational diagram is an empirical argument justifying the assumption of LTE emission. In instances where a single excitation temperature is unable to describe the data, either a temperature gradient or non-LTE processes should be taken into account. In the case of non-LTE effects, they often lead to a sub-thermal population of lines, implying that our mass estimate would serve as a lower limit, and the actual column density could be higher. However, taking into consideration the previous discussion on the low critical density for rotational H$_2$ emission, lower than the typical densities in warm molecular layers, the effect of a temperature gradient in the emitting gas would be a more likely explanation for a non-linear trend in the rotational diagram.

The observed H$_2$ emission traces only the innermost region of the disk and its mass represents only a small fraction of the total disk mass. The disk midplane, even at such small radii, could have temperatures lower than our estimate. Therefore, even in the inner disk, warm H$_2$ is a fraction of the total H$_2$ reservoir, and our mass estimate for the warm H$_2$ component is a lower limit to the total H$_2$ mass within the emitting radius. From the non-detection of millimeter continuum emission, \citet{Barenfeld16} give an upper limit to the disk mass of 0.2~M$_\mathrm{Jup}$. By making a few assumptions, we can give a rough estimate of the disk mass based on our measured warm H$_2$ mass. We assumed a small disk size $r_0 = 10$~au, and a power-law distribution for the disk surface density distribution:

\begin{equation}
    \Sigma(r) = \Sigma_0 \left( \frac{r}{r_0} \right)^{-1}.
\end{equation}

By knowing the gas surface density at $r = 0.033$~au from our H$_2$ emission analysis, we can use this equation to get a total disk mass estimate of 0.05~M$_\mathrm{Jup}$, which is in agreement with the upper limit of 0.2~M$_{Jup}$ from \citet{Barenfeld16}.

\subsection{Atomic hydrogen}
\label{sec: atomic hydrogen}
Atomic hydrogen recombination lines trace stellar accretion and ejection processes, as demonstrated by their broad profiles consistent with almost free-falling material \citep[e.g.,][]{Alcala14}. These lines are now commonly used to measure mass accretion rates in YSOs \citep{Antoniucci11, Antoniucci14, Biazzo12, Antoniucci16, Gravity23}, although they can also originate from winds and jets \citep{Beck10, Bally16, Ercolano17}. While it is necessary to account for both accretion and wind processes to reproduce all the spectral features, a tight empirical correlation has been measured between the integrated flux of HI recombination lines and the accretion luminosities $L_{acc}$. This correlation has been extensively studied for optical and near-infrared lines, such as the H$\alpha$, Br$\gamma$, Pa$\beta$ \citep{Antoniucci14, Alcala14, Alcala17}, and Pf$\beta$ (at 4.6~$\mu$m) lines \citep{Salyk13}. However, for the embedded phases of star formation or sources significantly affected by extinction, we must rely on longer wavelengths. Only a few space-based studies are available in the literature, based on data obtained with the Infrared Space Observatory and \textit{Spitzer} space telescope \citep[e.g.,][]{vanDishoeck98, Gibb00, Molinari08, An11}. In particular, \citet{Rigliaco15} explores the relations between the HI~(7-6) line luminosity at 12.37~$\mu$m and the accretion luminosity using \textit{Spitzer} data on a sample of classical T~Tauri stars. With the MIRI instrument, it is now possible to study in detail the mid-infrared atomic hydrogen recombination lines.

\begin{figure*}
    \centering
    \includegraphics[width=\linewidth]{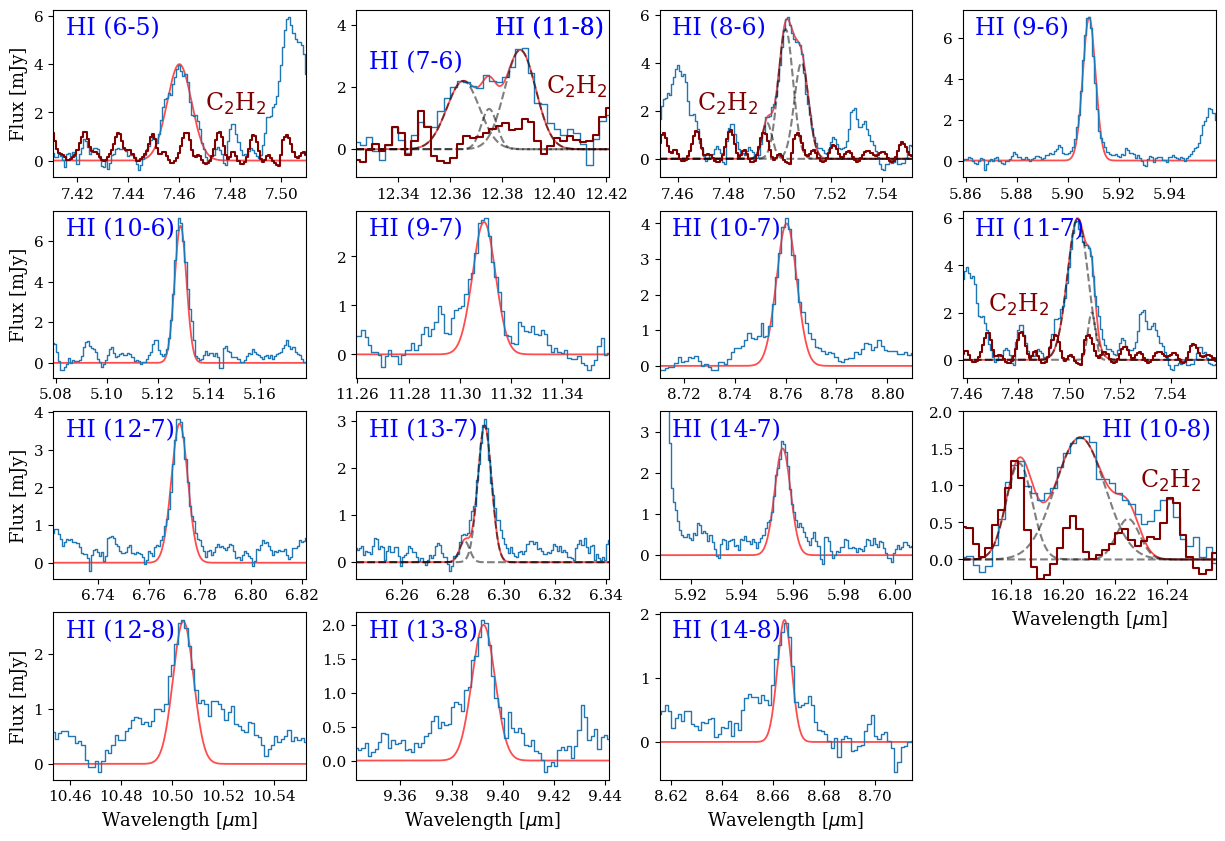}
    \caption{Detected atomic hydrogen recombination lines. The red profiles show line fitting using one or multiple Gaussian functions (dashed lines) in the case of line blending with other spectral features. The dark red profile is the C$_2$H$_2$ model from \citet{Tabone23}}
    \label{fig: HI}
\end{figure*}

\begin{figure}
    \centering
    \includegraphics[width=\linewidth]{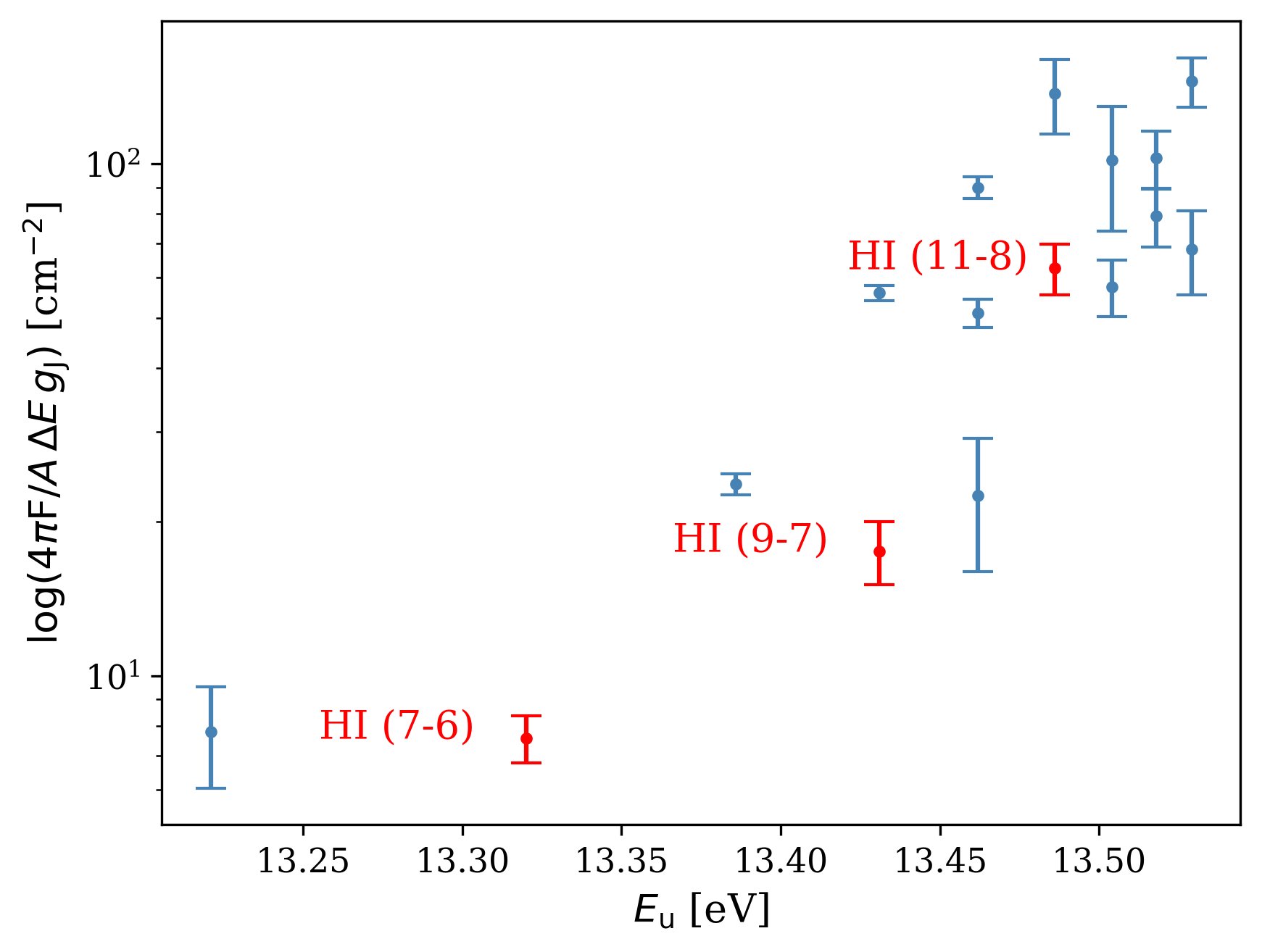}
    \caption{Excitation diagram of the HI lines. These lines are produced by shocked material for which the Boltzmann distribution does not apply, therefore the y-axis is in terms of integrated line flux rather than column density. In red, we highlight the lines used for the comparison to atomic line emission models.}
    \label{fig: HI rot diagram}
\end{figure}

We identified 16 atomic hydrogen recombination lines in the spectrum of J160532, with the principal quantum number of the lower energy levels between 5 and 8 (Pfund, Humphreys, and higher series). To estimate the line fluxes, we followed the same procedure used to analyze H$_2$ in Sec.~\ref{sec: molecular hydrogen}, and the lines are shown in Fig.~\ref{fig: HI}. The main Gaussian component of the line fits has a width ranging from about 100 to 200~km~s$^{-1}$, compatible with the local spectral resolution. Our results are summarized in Tab.~\ref{tab: HI}. The line fluxes are shown in an excitation diagram in Fig.~\ref{fig: HI rot diagram}. This diagram highlights that the intensities of the HI lines deviate from what would be expected in optically thin emission conditions. Moreover, lines with the same upper energy level do not overlap in the plot, contrary to what would be expected in the scenario of optically thin emission.

\begin{table*}[]
    \caption{Measured parameters of the identified atomic hydrogen recombination lines.}
    \label{tab: HI}
    \centering
    \begin{tabular}{c  c c}
        \hline
        \hline
        Transition & Line Center & Integrated Flux\\
         & [$\mu$m] & $10^{-16}$ [erg~s$^{-1}$~cm$^{-2}$]\\
        \hline
        6-5 & 7.460 & $25.9 \pm 5.8$\\
        7-6 & 12.372 & $9.4 \pm 1.0$\\
        8-6 & 7.503 & $20.9 \pm 1.0$\\
        9-6 & 5.908 & $37.7 \pm 1.3$\\
        10-6 & 5.129 & $44.6 \pm 2.2$\\
        9-7 & 11.309 & $7.1 \pm 1.0$\\
        10-7 & 8.760 & $15.7 \pm 1.0$\\
        11-7 & 7.508 & $32.0 \pm 5.3$\\
        12-7 & 6.772 & $19.4 \pm 5.3$\\
        13-7 & 6.292 & $14.9 \pm 1.9$\\
        14-7 & 5.957 & $17.1 \pm 1.9$\\
        10-8 & 16.209 & $4.5 \pm 1.3$\\
        11-8 & 12.387 & $9.7 \pm 1.1$\\
        12-8 & 10.504 & $7.1 \pm 0.9$\\
        13-8 & 9.392 & $7.7 \pm 1.0$\\
        14-8 & 8.665 & $5.3 \pm 1.0$\\
        \hline
    \end{tabular}
\end{table*}

We measured the mass accretion rate from the HI~(7-6) line at 12.37~$\mu$m  using the method introduced by \cite{Rigliaco15}, who derives the following empirical relation between the HI~(7-6) integrated flux and the accretion luminosity in classical T~Tauri stars:

\begin{equation}
    \label{eq: rigliaco}
    \log L_{\mathrm{HI}~(7-6)} / L_\odot = \left( 0.48 \pm 0.09 \right) \times \log L_{acc} / L_\odot - \left( 4.68 \pm 0.10 \right).
\end{equation}

\noindent The standard relation between the mass accretion rate $\dot{M}_{acc}$ and $L_{acc}$ was introduced by \cite{Gullbring1998}:

\begin{equation}
    \label{eq: mass accretion rate}
    \dot{M}_\mathrm{acc} = \frac{L_\mathrm{acc} R_\star}{G M_\star} \left( 1 - \frac{R_\star}{R_\mathrm{in}} \right)^{-1},
\end{equation}

\noindent where $M_\star = 0.14$~M$_\odot$ is the stellar mass, $R_\mathrm{in}$ the radius at which the stellar magnetosphere disperses the accreting gas, and $R_\star = 0.45$~R$_\odot$ is the stellar radius. $R_\mathrm{in}$ is usually assumed to be $5 \, R_\star$ \citep{Herczeg08, Alcala14, Rigliaco15}, while $R_\star$ is derived from the stellar effective temperature $T_\mathrm{eff} = 3850$~K and the stellar luminosity $L_\star$ using the Stefan–Boltzmann relation $L_\star = 4 \pi R_\star^2 \sigma T_\mathrm{eff}^4$, where $\sigma$ is the Stefan–Boltzmann constant and $T_\mathrm{eff}$ is estimated from the spectral type M4.75 \citep{Fang17}.

\begin{figure}
    \centering
    \includegraphics[width=\linewidth]{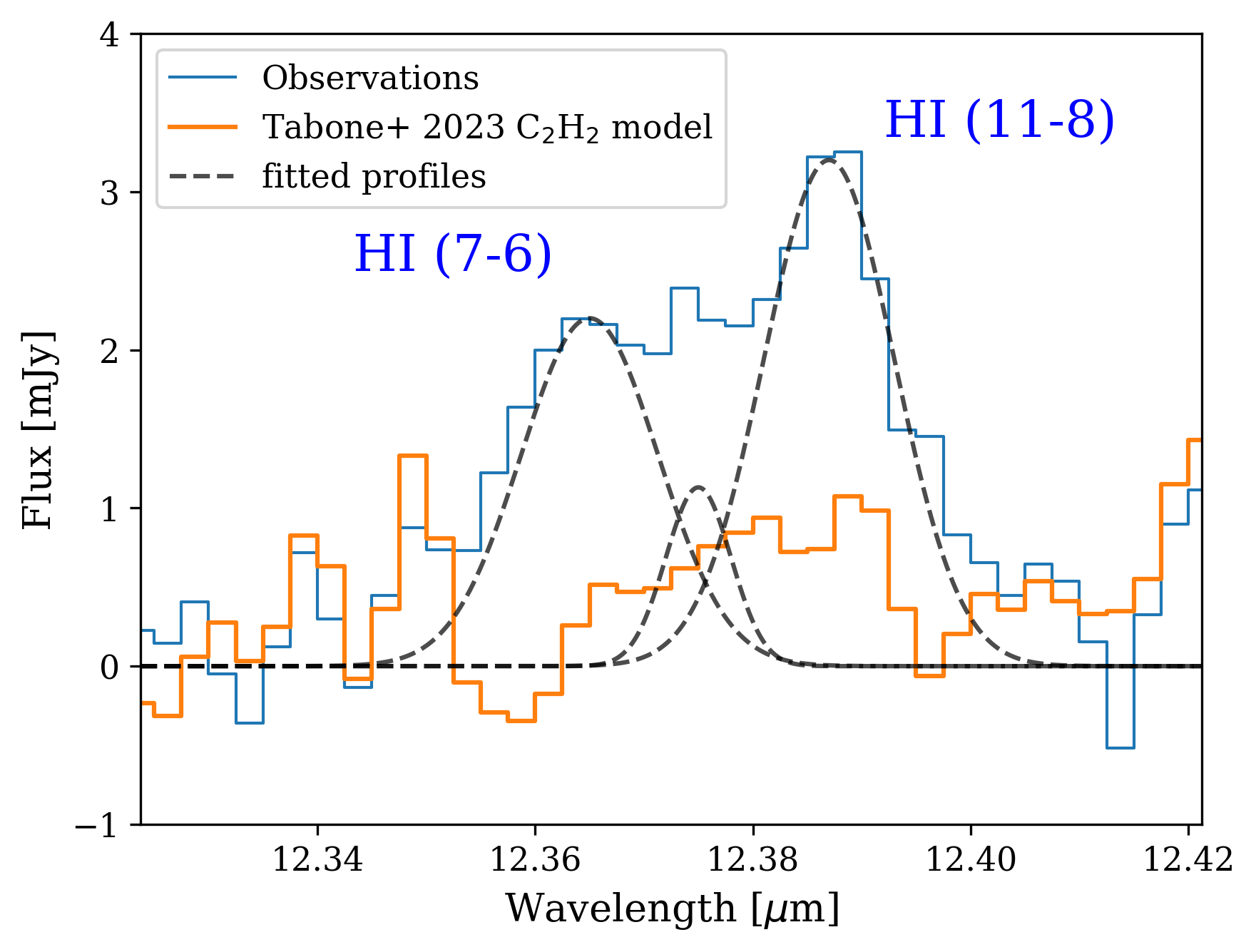}
    \caption{Emission lines used to measure the accretion rates. The black dashed profile on the left is the HI~(7-6) line, the one on the right is the HI~(11-8) line, while the central component is C$_2$H$_2$ emission, according to the predictions of \citet{Tabone23} (the orange profile).}
    \label{fig: accretion lines}
\end{figure}

Notably, the HI~(7-6) line at 12.372~$\mu$m is located near another atomic hydrogen recombination line, the HI~(11-8) transition at 12.387~$\mu$m. This is evident from the observed line profile (see Fig.~\ref{fig: accretion lines}), which displays two dominant Gaussian components, each approximately 150~km~s$^{-1}$ wide. The HI~(7-6) line is shifted by about -4~km~s$^{-1}$ compared to its rest frequency, while the HI~(11-8) by about 64~km~s$^{-1}$, which is compatible with the line rest frequency given the local resolution of about 110~km~s$^{-1}$. Additionally a third, weaker component is present, with a width similar to the local resolution element, around $\sim 60$~km~s$^{-1}$. Comparatively, the widths of the two main components align with those of the other atomic hydrogen lines. We identified the component at the shorter wavelength as the HI~(7-6) line, while the component at the longer wavelength corresponds to the HI~(11-8) line. The third component could be the $Q$-branch of a weak $P$, $Q$, and $R$-branch pattern of optically thick C$_2$H$_2$ emission modeled by \citet{Tabone23}, centered at 12.379~$\mu$m (Fig.~\ref{fig: accretion lines}). The two identified components have peak flux densities of 2.3~mJy and 3.5~mJy, and luminosities of $4.9 \times 10^{-7}$~L$_\odot$ and $6.8 \times 10^{-7}$~L$_\odot$ respectively. From the HI~(7-6) luminosity we get an accretion luminosity of $4.0 \pm 2.5 \times 10^{-4}$~L$_\odot$ using Eq.~\ref{eq: rigliaco}, corresponding to a mass accretion rate of $1.0 \times 10^{-10}$~M$_\odot$~yr$^{-1}$. However, this relation was calibrated by \citet{Rigliaco15} using \textit{Spitzer} data, which lacked the spectral resolution necessary to distinguish between the HI(7-6) and HI(11-8) profiles. Since both lines are excited at similar energies, the excitation of one line implies the excitation of the other. Therefore, we argue that the fluxes used by \citet{Rigliaco15} to measure the accretion luminosity are not solely the HI~(7-6) line but rather the combined fluxes of both the HI~(7-6) and HI~(11-8) lines. The contribution to the total flux given by the C$_2$H$_2$ component is negligible and does not affect the measured accretion luminosity. By considering the total line flux, we obtained an accretion luminosity of $(3.1 \pm 1.9) \times 10^{-3}$~L$_\odot$, corresponding to a mass accretion rate of $(4.0 \pm 2.5) \times 10^{-10}$~M$_\odot$~yr$^{-1}$. This result is in good agreement with the mass accretion rate estimate of $4.17 \times 10^{-10}$~M$_\odot$~yr$^{-1}$ reported by \citet{Fang23} based on the H$\alpha$ emission. A follow-up study utilizing the higher spectral resolution data from the JWST on a sample of sources could validate the hypothesis that the accretion rate measured using Eq.~\ref{eq: rigliaco} uses the combined HI~(7-6) and HI(11-8) fluxes.

Finally, \citet{Antoniucci16} study the Balmer and Paschen decrements in a sample of 36 low-mass Class~II sources. While they do not find any correlation between the line decrements and the source properties, they find a tentative correlation between the decrement shape and the mass accretion rate. If we assume their results to be valid also for other HI series (a reasonable assumption since they originate from the same physical process), the decrement shape in our source (Fig.~\ref{fig: decrements}) falls in their Type 3, bumpy profile category. The disks in this category have mass accretion rates between $10^{-10} - 10^{-9}$~M$_\odot$~yr$^{-1}$, which is in accord with our measured value of $(4.0 \pm 2.5) \times 10^{-10}$~M$_\odot$~yr$^{-1}$.

\subsection{Atomic line emission models}
To constrain the physical properties of the gas from which the  HI lines originate, we compared the observed line fluxes to predictions from two standard emission models. The first model employed was the classic Case~B recombination line model \citep{Baker38, Storey95}, which assumes that Lyman lines are optically thick and all other transitions are optically thin, and that level populations are determined through radiative cascade from the continuum. The second model utilized is the more recent local line excitation approach developed by \citet{Kwan11}, which considers the local physical conditions of the gas to consistently evaluate the line emissivities. The input physical parameters of this model are the gas temperature, the hydrogen density, the ionization rate (which was not specified in the Case~B models), and the gas velocity gradient transverse to the radial direction. These parameters are chosen to reflect the typical physical condition in accreting YSOs.

\begin{figure}
    \centering
    \includegraphics[width=\linewidth]{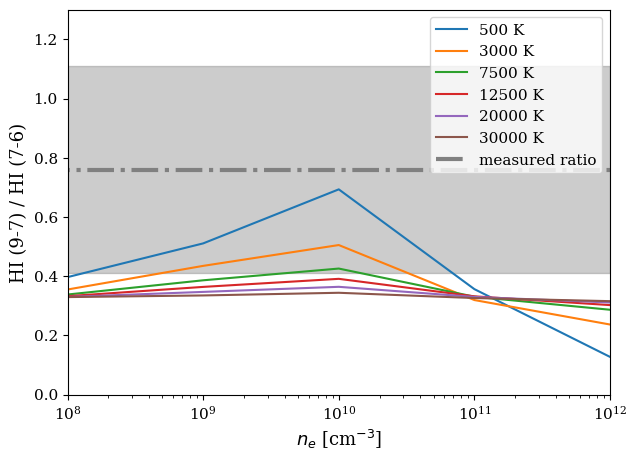}
    \caption{HI~(9-7) and HI~(7-6) line ratio as a function of electron density from the Case B recombination model compared to our measured ratio (gray dotted line). The shaded area is the uncertainty of the measured ratio.}
    \label{fig: Case B}
\end{figure}

The Case B recombination model is commonly used in the literature to derive the physical properties of accreting gas onto T~Tauri stars. It assumes that the gas is optically thick to Lyman series photons and optically thin to the photons linked to other transitions. \citet{Storey95} computed the line fluxes for hydrogenic atoms as a function of temperature and electron density for the Case B recombination model, which are available online\footnote{http://cdsarc.u-strasbg.fr/viz-bin/Cat?VI/64}. Using this model, we compare in Fig.~\ref{fig: Case B} the HI~(9-7) and HI~(7-6) lines for a temperature range of 500-30\,000~K and electron density $n_e$ between $10^8$~cm$^{-3}$ and $10^{12}$~cm$^{-3}$ to our measured value of $\sim 0.76 \pm 0.35$. According to this model, the gas temperature varies widely and it is poorly constrained, ranging from 500~K to 12\,500~K, electron density between $10^8$~cm$^{-3}$ and $10^{11}$~cm$^{-3}$. The highest temperatures are observed at column densities of $\sim 10^{10}$~cm$^{-3}$. 

While we detected numerous HI lines, our analysis centers on the HI~(9-7), (7-6), and (11-8) transitions. The physical properties of the emitting gas can be probed using any line of choice. These bright lines are favored due to observational limitations, enabling us to effectively constrain gas properties using only a subset of observed lines. Although a broader selection of lines would provide tighter constraints, our options are constrained by available atomic line emission models and our spectral window, limiting us to the HI~(9-7), (7-6), and (11-8) transitions that have been previously modeled \citep{Storey95, Kwan11}. Our study highlights how the JWST-MIRI instrument expands the range of observable lines, leading to a more comprehensive understanding of protoplanetary disk properties through improved modeling efforts with this new data.

\begin{figure}
    \centering
    \includegraphics[width=0.8\linewidth]{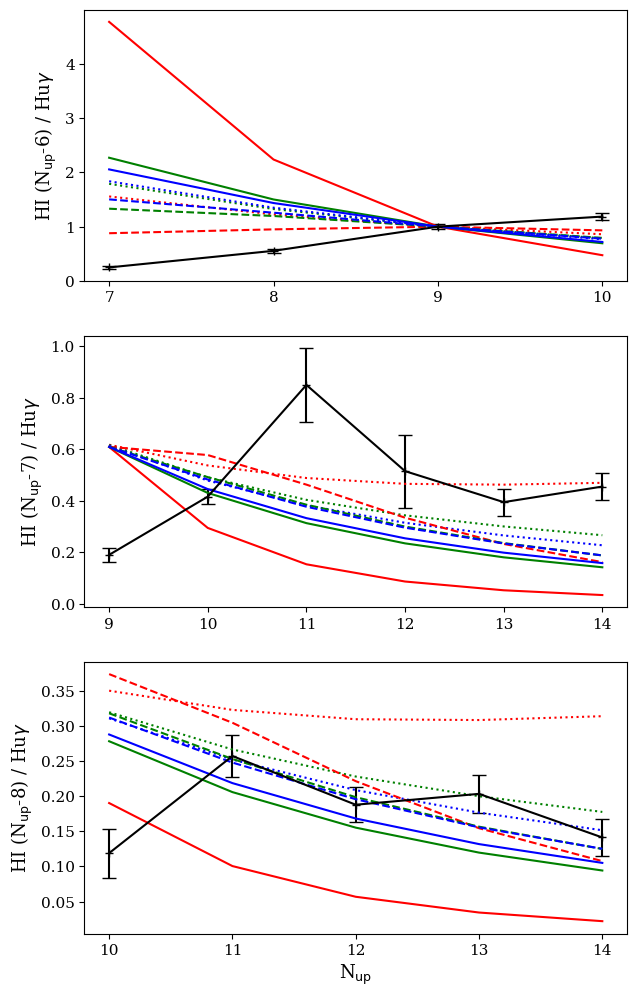}
    \caption{Comparison of the three detected line series to the Case~B model predictions. The solid black lines show the observed line fluxes for transition with lower energy levels (from top to bottom) 6, 7, and 8 relative to the reference line HI~(9-6). The lines with different colors are the predicted values by the Case~B model for different densities and temperatures of the emitting gas. The model densities are $10^{8}$~cm$^{-3}$ (dotted line), $10^{10}$~cm$^{-3}$ (slashed line), and $10^{12}$~cm$^{-3}$. The temperatures are 500~K (red), $5\,000$~K (green), and $10\,000$~K (blue).}
    \label{fig: decrements}
\end{figure}

In Fig.~\ref{fig: decrements} we show the line ratios as a function of the upper level quantum number $N_\mathrm{up}$, for different line series, with lower energy levels 6, 7, and 8. In computing the ratio, we used as reference the HI~(9-6) (Hu$\gamma$) line, the cleanest line in our sample. However, choosing a different line does not significantly affect the shape of the line ratio distribution. In Fig.~\ref{fig: decrements} we also show the predicted values for the Case~B model for a range of temperatures from 500~K to $10\,000$~K and densities from $10^8$~cm$^{-3}$ to $10^{12}$~cm$^{-3}$. The Case~B line ratios poorly reproduce the observations, especially for the lower lines in the series. This plot highlights the effect of the optical depth on the line fluxes. In the case of optically thin emission, we expect the ratio of lines with the same upper level to be the ratio of the Einstein coefficients of the transitions. However, our observed lines contradict this expectation. For instance, the ratio of the HI~(9-7) and HI~(9-6) lines is about 5.3. This value is much higher than the Einstein coefficient ratio of about 0.9. This discrepancy suggests that these lines cannot be considered optically thin, as is assumed in the Case~B recombination theory. This model therefore is not an appropriate description of these data.

\begin{figure}
    \centering
    \includegraphics[width=\linewidth]{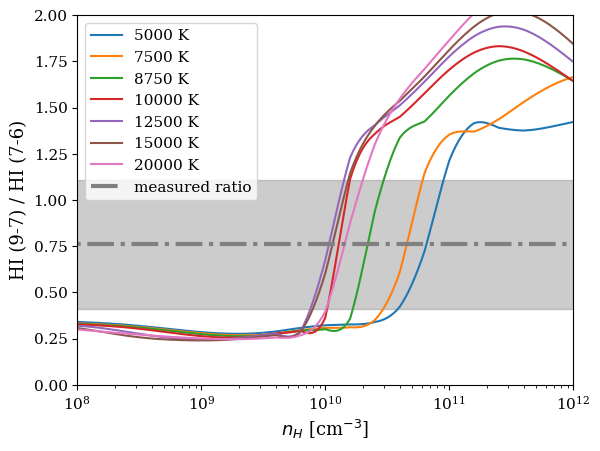}
    \caption{HI~(9-7) and HI~(7-6) line ratios as a function of atomic hydrogen density according to the KF atomic emission model \citep{Kwan11} compared to our measured ratio, the gray dotted line. The shaded area is the uncertainty of the measured ratio.}
    \label{fig: KD}
\end{figure}

The HI~(9-7) and HI~(7-6) line ratio was also computed following a different approach by \citet{Kwan11}, hereafter denoted KF model, where they measure the optical depth of each transition instead of assuming optically thin emission. They also included in their calculation the gas ionization rate and the velocity gradient, not accounted for in the Case~B recombination model and tuned these parameters to reflect the typical physical condition of accreting YSOs. Using the data available online\footnote{http://iopscience.iop.org/0004-637X/778/2/148/suppdata/data.tar.gz} we compared again the model ratio of the HI~(9-7) and HI~(7-6) line to our measurement, for temperatures between 5\,000~K and 20\,000~K and atomic hydrogen densities between $10^8$~cm$^{-3}$ and $10^{12}$~cm$^{-3}$ (see Fig.~\ref{fig: KD}). The KF model predicts that all the temperatures are compatible with our measured line ratios, and constrains the hydrogen density to $7 \times 10^9 - 4 \times 10^{10}$~cm$^{-3}$. Compared to the Case~B model, a proper treatment of the lines optical depth gives us much tighter constraints. Since the ionization degree of $n_e/n_H$ typically ranges between 0.1-1.0, this density range is in agreement with the constraints from the Case B model. Case~B models with temperatures up to $7\,500$~K and atomic hydrogen densities of about $10^{10}-10^{11}$~cm$^{-3}$ (with $n_e / n_H = 0.1$) are in agreement with the low temperature (5\,000~K), high density (a few $10^{10}$~cm$^{-3}$) KF models, while densities below $10^{10}$~cm$^{-3}$ cannot be reached given our measured line ratio. However, we could not give an upper limit to the atomic hydrogen densities according to the KF models, as the line ratio was not calculated for temperatures lower than 5\,000~K. When looking at the model prediction of the other HI line fluxes of the Case~B model, the fit to the data is quite poor. The observed line fluxes require temperatures lower than 500~K. These low temperatures are unlikely for shocked accreting HI gas. This indicates that these lines may not be optically thin, as assumed in the Case~B model \citep{vanderAncker00, Edwards13, Rigliaco15}, and this model cannot be used to model the emission of high-density regions such as accretion flows in YSOs. 

\section{Discussion}
H$_2$ emission in classical T~Tauri stars can originate from either the inner region of a protoplanetary disk, excited by UV radiation, or shocked gas in stellar outflows \citep{Beck10, Carmona11}. We analyzed the profiles of the detected H$_2$ lines to constrain the origin and formation mechanism of these lines. As discussed in Sec.~\ref{sec: molecular hydrogen}, the H$_2$ emission comes from a spatially unresolved component with no evidence of outflows. The emission is observed at a velocity consistent with the rest velocity of the star. However, the data are also compatible with a stellar wind or outflow coming from the very inner disk with a projected velocity $\lesssim 90$~km~s$^{-1}$. To further constrain the origin of these lines, we measured the full width at zero intensity (FWZI) of the profiles, giving us the maximum projected velocity of the emitting H$_2$, and compared this value to the Keplerian velocity at the emitting radius 0.033~au. We measured a mean FWZI of $160 \pm 110$~km~s$^{-1}$ and given that the Keplerian velocity at the emitting radius is 60~km~s$^{-1}$, this measurement is compatible with Keplerian motions of the gas within the assumed emitting radius. However, it is important to note that we lack information about the disk geometry, and the projected Keplerian velocity might be below 60~km~s$^{-1}$, depending on the disk inclination.

We performed the same analysis of the FWZI of the HI line profiles to investigate their origin. As discussed in Sec.~\ref{sec: atomic hydrogen}, we assumed that HI emission traces accretion flows of free-falling gas on the stellar surface. Following the assumption that the gas is accreting from a radius $R_{in} = 5 \, R_\star$ (see Sec.~\ref{sec: atomic hydrogen}), we got a free-fall velocity of 315~km~s$^{-1}$. The average FWZI for HI lines is $370 \pm 140$~km~s$^{-1}$, supporting the hypothesis that this emission comes from free-falling gas accreting on the central object.

The accretion rate we derived using the atomic hydrogen lines is consistent with those measured in other YSOs. \citet{Alcala17} measured the accretion rate of YSOs in the Lupus region. Their measured accretion rates range from $\sim 5 \times 10^{-12} \; M_\odot$~yr$^{-1}$ to $\sim 6 \times 10^{-8} \; M_\odot$~yr$^{-1}$, and our measured mass accretion rate of $\sim 4 \times 10^{-10} \; M_\odot$~yr$^{-1}$ correspond to an average accretion for the YSOs in their sample. \citet{Alcala17} also derive an analytical relation between $L_{acc} - L_\star$ and one between $\dot{M}_{acc} - M_\star$. Using their relation for $L_{acc}$:

\begin{equation}
    \label{eq: lacc from lstar}
    \log L_{acc} = (1.26 \pm 0.14) \cdot \log L_\star - (1.60 \pm 0.13),
\end{equation}

\noindent we found $L_{acc} = (4.4 \pm 2.4) \times 10^{-4} \; L_\odot$. This measurement is about one order of magnitude lower than our measured value $(3.1 \pm 1.9 )\times 10^{-3} \; L_\odot$. There is, however, a significant scatter in the data used to derive Eq.~\ref{eq: lacc from lstar}, and our measurement falls within this scatter. Using their relation for $\dot{M}_{acc}$ for low-mass stars ($< 0.2 M_\odot$):

\begin{equation}
    \log \dot{M}_{acc} = (4.58 \pm 0.68) \cdot \log M_\star - (6.11 \pm 0.61),
\end{equation}

\noindent we derived $M_\mathrm{acc} = (1.0 \pm 1.8) \times 10^{-10} \; M_\odot$~yr$^{-1}$, in agreement with our measured value of $(4.0 \pm 2.5) \times 10^{-10}$. 

Numerous spectroscopic surveys in several star formation regions have observed a correlation between the accretion rate $\dot{M}_{acc}$ and the stellar mass $M_\star$ \citep{Alcala14, Venuti14, Venuti19, Manara16, Manara17, Hartmann16}, with reported slopes ranging from approximately 1.6 to 2. \citet{Manara23} compare these data and presented their findings in their Fig.~4. Their analysis demonstrates that the observations align well with the relationship $\dot{M}_{acc} \propto M_\star^2$. The accretion rate for this particular source adheres to the $\dot{M}_{acc} \propto M_\star^2$ relationship, classifying it as a standard accretor compared to other YSOs. It is important to highlight that our determined accretion rate is not exceptionally high, despite the prominent presence of HI lines in the spectrum. In fact, it falls within the lower range when compared to similar sources with measured accretion rates \citep[e.g.,][]{Rigliaco15, Manara16, Alcala17}. This clear detection of HI lines is likely attributed to the remarkably low dust extinction, enabling us to penetrate deeply into the disk and obtain an unobstructed view of the accretion layer.



\section{Conclusions}
In this paper, we present the full JWST-MIRI spectrum of the disk around the very low-mass star 2MASS-J16053215-1933159. These observations reveal numerous atomic and molecular hydrogen lines in the mid-infrared spectrum of this source. We utilize HI recombination lines to measure the accretion rate of the source and H$_2$ emission to constrain the rotational temperature and column density of the disk gas.

With the high resolving power and sensitivity of the MIRI-MRS instrument, we detect several pure rotational $\nu = 0-0$ H$_2$ transitions. The observed lines are optically thin due to their small transition probabilities, allowing us to constrain the temperature and column density of the gas. To accurately measure the line fluxes, we performed a linear fit of the continuum and fit Gaussian profiles to the H$_2$ line profiles, while accounting for blending with other spectral features. Using the measured fluxes, we constructed a rotational diagram to estimate the temperature and column density of the emitting gas. Under the assumption of optically thin emission in local thermodynamic equilibrium, our calculations yield a total mass of warm H$_2$ gas at $(2.3 \pm 1.3) \times 10^{-5}$~M$_\mathrm{Jup}$ with a corresponding temperature of $635 \pm 94$K. This represents only a minor portion of the entire disk mass, which is constrained to be below 0.2~M$_\mathrm{Jup}$ due to the upper limit given by the non-detection of millimeter continuum emission.

We investigate the HI~(7-6) line profile, a tracer for mass accretion. Through analyzing its profiles, we identify two main Gaussian components corresponding to the HI~(7-6) and HI~(11-8) atomic hydrogen recombination lines. These lines are in close proximity in wavelength, and the flux measurement used by \citet{Rigliaco15} to assess accretion luminosity likely included contributions from both lines, rather than solely the HI~(7-6) flux. When considering the combined flux, our estimated accretion luminosity $4.17 \times 10^{-10}$~M$_\odot$yr$^{-1}$ is consistent with the measurement based on H$\alpha$ emission by \citet{Fang23}. This suggests that the previous calibration of accretion luminosities may include the contribution from both lines. To further validate our hypothesis, future observations with higher spectral resolution than \textit{Spitzer} data, such as those possible with the JWST, would be valuable. We compare our measured accretion rate with other YSOs, finding that this source has typical accretion properties.

To derive the physical properties of the accreting gas onto the star, we compare the HI~(9-7) to HI~(7-6) line ratio with the Case~B recombination model \citep{Baker38, Storey95} and the Kwan \& Fischer (KF) model \citep{Kwan11}. The Case~B model provides limited constraints on the density and temperature, primarily because the lines are not optically thin, as assumed by this model. The KF model predicts that our measured line ratios are compatible with a range of temperatures from 5\,000~K to 20\,000~K, and atomic hydrogen densities between 10$^9$~cm$^{-3}$ and 10$^{11}$~cm$^{-3}$. The atomic hydrogen density is well constrained by the KF~model, while the temperature is not. However, the Case~B model fails to reproduce the observed line fluxes and points to temperatures lower than 500~K, not physical for accretion flows.

In summary, our observation and analysis of one of the first JWST MIRI MRS data of a faint disk reveal new important results on the circumstellar gas properties. We observed several H$_2$ rotational lines, which constrain the warm emitting gas temperature and mass. Moreover, the measurement of HI recombination lines can be used to estimate the mass accretion rate onto the central source, while the comparison with theoretical models gives us information on the physical condition of the emitting gas. These results, while interesting on their own, demonstrate how further investigations of protoplanetary disks with JWST will dramatically enhance our understanding of these complex astronomical environments.


\begin{acknowledgements}
The following National and International Funding Agencies funded and supported the MIRI development: NASA; ESA; Belgian Science Policy Office (BELSPO); Centre Nationale d’Etudes Spatiales (CNES); Danish National Space Centre; Deutsches Zentrum fur Luftund Raumfahrt (DLR); Enterprise Ireland; Ministerio De Econom\'ia y Competividad; Netherlands Research School for Astronomy (NOVA); Netherlands Organisation for Scientific Research (NWO); Science and Technology Facilities Council; Swiss Space Office; Swedish National Space Agency; and UK Space Agency.

R.F., T.H., and K.S. acknowledge support from the European Research Council under the Horizon 2020 Framework Program via the ERC Advanced Grant Origins 83 24 28.

B.T. is a Laureate of the Paris Region fellowship program, which is supported by the Ile-de-France Region and has received funding under the Horizon 2020 innovation framework program and Marie Sklodowska-Curie grant agreement No. 945298.

A.C.G. has been supported by PRIN-INAF MAIN-STREAM 2017 “Protoplanetary disks seen through the eyes of new- genera-tion instruments” and from PRIN-INAF 2019 “Spectroscopically tracing the disk dispersal evolution (STRADE)”

G.B. thanks the Deutsche Forschungsgemeinschaft (DFG) - grant 138 325594231, FOR 2634/2.

E.v.D. acknowledges support from the ERC grant 101019751 MOLDISK and the Danish National Research Foundation through the Center of Excellence ``InterCat'' (DNRF150). 

I.K., A.M.A., and E.v.D. acknowledge support from grant TOP-1 614.001.751 from the Dutch Research Council (NWO).

I.K. acknowledge funding from H2020-MSCA-ITN-2019, grant no. 860470 (CHAMELEON).

O.A. and V.C. acknowledge funding from the Belgian F.R.S.-FNRS.

O.A, V.C, and D.G. thank the Belgian Federal Science Policy Office (BELSPO) for the provision of financial support in the framework of the PRODEX Programme of the European Space Agency (ESA).

D.R.L. acknowledges support from Science Foundation Ireland (grant number 21/PATH-S/9339).

M.T. acknowledges support from the ERC grant 101019751 MOLDISK.

\end{acknowledgements}

%
%

\bibliographystyle{aa} 
\bibliography{aanda} 

\end{document}